# Drift Diffusion and energy-transport in degenerate hopping systems

Dan Mendels and Nir Tessler

The Sarah and Moshe Zisapel nanoelectronic center, Electrical Engineering Dept., Technion Israel institute of technology, Haifa 32000, Israel

Abstract

Revisiting charge transport in degenerate semiconductors we present a modification to the drift diffusion equation where instead of employing the generalized Einstein relation we add an energy flux term thus solving several inconsistencies. This leads also to the conclusion that, contrary to common belief, a constant quasi-Fermi level does not necessarily ensure zero electrical current. While we revisit the drift diffusion process in the context of degenerate hopping systems, a considerable part of the argumentation put forward can be applied generally to degenerate systems.

***Introduction*** – Understanding the underlying physics of modern devices often calls for an ever increasing details of their modeling. For example, hopping transport in disordered semiconductors may exhibit rich phenomena as dispersive transport in either time[1] or space[2], electric field dependence[3-9], and charge density dependence[10-16]. However, large body of the experimental analysis relies on the drift diffusion equation and hence we reexamine its completeness. A reduction of the Boltzmann transport equation (BTE) results in the carrier continuity and energy balance equations that describe the flow of charge carriers ($J$) and of energy ($S$) (see for example chapter 3.9 in [17]). It has been shown that for non-localized systems under non degenerate conditions the energy balance is dependent only on temperature gradients and hence, in the absence of thermal gradients, it is common to use only the basic and simple drift diffusion equation for the charge carrier flow:

(1) $J = qD\frac{dn}{dx} + qn\mu F$ .

While this equation is indeed highly robust and useful in analyzing devices it has become clear that in order to better describe the transport and through that be able to correlate between transport properties of different materials embedded within different device geometries one has to “correct” this equation[18]. For example, in the context of hopping transport in a Gaussian density of states (DOS), it was initially modified by letting the mobility be electric field dependent, charge density dependent and later by letting the ratio between mobility and diffusion follow the generalized Einstein relation. In the context of devices, although the Einstein relation has very little effect on the diode ideality factor[19] it still affects the charge density distribution, hence it received substantial attention. However, it seems that these modifications are not sufficient to capture the transport phenomena across a range of material and device parameters and efforts have been made to farther correct the above dependencies by finding the “true” DOS[20-22]. In this paper we examine whether part of the inconsistency found may be related to the fact that under most practical experimental conditions the semiconductors of interest are degenerate. We find that for degenerate hopping systems the use of equation (1) leads to the mobility as well as the Einstein relation being dependent also on the charge density gradient. The physical origin is that for the degenerate case the average energy of the charge carriers is dependent on the charge density suggesting that neglecting the energy balance equation describing the energy flow ($S$) in the system may be too crude of an approximation for a degenerate system.

***Methodology*** - To analyze the transport in a close to realistic amorphous organic semiconductor we assume the DOS to be Gaussian. To enhance the reliability of the method and gain physical insight we employ two methods of evaluation: the Monte Carlo simulation and the effective medium approximation (EMA). To be able to use the EMA to extract farther physical insight in the current context, we compare data obtained by its employment with data obtained via Monte Carlo simulations implemented under the same physical conditions. To this end we also limit the study to a low disorder case ($\sigma$=3kT) and make sure to employ relatively low electric fields ($F$<$10^4$V/cm) and not too low charge densities (n>$10^{16}$cm$^{-3}$) so as to minimize the effects associated with dispersive transport and the formation of conduction paths[2, 7, 23, 24]. The EMA formalism we chose is incorporated with the transport-energy concept which is calibrated to uniform carrier density Monte Carlo simulations.

When employing an approximate approach, as the EMA,[9] the total current density ($J$) between two parallel neighboring lattice planes i and j in the effective medium, placed along a given applied field (F) vector, can be expressed as $J = qn_i a W^+{}_e - qn_j a W_e^-$. Where $W_e^+$ and $W_e^-$ are the effective hopping rates between sites in the direction of the field and against it, respectively. $n_i$ and $n_j$ are the carrier densities in the two referred planes, q is the carrier charge, and 'a' is the distance between the planes. This relation can also be presented in a form similar to equation (1):

(2) $J = q\frac{\left(W_e^+ + W_e^-\right)}{2}a^2\frac{\left(n_i - n_j\right)}{a} - q\left(W_e^+ - W_e^-\right)a\frac{\left(n_i + n_j\right)}{2} = qD_1\frac{\left(n_i - n_j\right)}{a} + q\mu_1 F\frac{\left(n_i + n_j\right)}{2}$

We used these equations to express, through the hopping rates, the diffusion coefficient ($D_1$) and the mobility value ($\mu_1$), the latter being valid only for nonzero electric field ($F$). The above representation of D and $\mu$ is one of the three to be discussed in this paper and hence the subscript 1. Here it should be noted that the above definitions coincide with the analogous expressions obtained when solving an a-symmetric random walk problem where the hopping probabilities are proportional to $W_e^+$ and $W_e^-$.

The results presented below rely on the implementation of the effective rates ( $W_e$ ) using the Effective Medium Approximation (EMA), developed in [9] for low disorder systems and later extended for higher disorder systems, using the concept of transport energy ($E_t$) in [25, 26]:

$$(3)\begin{cases} W^+{}_e = \dfrac{\int_{-\infty}^{E_t} g(E_i)\, f(E_i, E_{Fi})\, \nu_0 \exp\left(\dfrac{-2a}{b} - \dfrac{\left|E_t - (E_i - qFa)\right| + \left(E_t - (E_i - qFa)\right)}{2K_BT}\right) dE_i}{\int_{-\infty}^{E_t} g(E_i)\, f(E_i, E_{Fi})\, dE_i} \\ W^-{}_e = \dfrac{\int_{-\infty}^{E_t} g(E_j)\, f(E_j, E_{Fj})\, \nu_0 \exp\left(\dfrac{-2a}{b} - \dfrac{\left|E_t - E_j\right| + \left(E_t - E_j\right)}{2K_BT}\right) dE_j}{\int_{-\infty}^{E_t} g(E_j)\, f(E_j, E_{Fj})\, dE_j} \end{cases}$$

Here the integrals are applied to the product of the system's DOS in lattice plane i,

$g(E_i) = N_0 \frac{1}{\sqrt{2\pi\sigma^2}} e^{\frac{-E_i^2}{2\sigma^2}}$ , the Fermi-Dirac distribution function

$f_{FD}(E_i, E_{Fi}) = \frac{1}{1 + \exp\left(\frac{E_i - E_{Fi}}{K_BT}\right)}$ and the Miller-Abrahams expression for hopping

from a site affiliated with an energy $E_i$ to a site affiliated with the transport energy $E_t$. $E_{Fi}$ is the quasi-Fermi level. Also in (3) 'a' is the lattice constant, b the localization length of the electronic states, F the applied field, $K_B$ the Boltzmann constant and T the system's temperature. Note that due to $E_t$ being the target energy in both directions in (3), qFa appears only in the forward jump rate to ensure that the effect of the electric field on the ratio between forward and backward jumps is $\exp\left(\frac{qFa}{K_BT}\right)$. When using the Monte Carlo simulations one needs to avoid artifacts in the extraction of *D* and $\mu$ hence we followed the methodology presented in [24].

To place the semi-analytical formulation's [eq. (2)] calculations and the Monte Carlo simulations on equal footing we employed the following procedure. First, we used the Monte Carlo simulations to calculate the charge carrier mobility as a function of charge density using cyclic boundary conditions in all three directions (i.e. homogenous carrier distribution). This type of calculation has been reported in the past and indeed the results reported in Figure 1 (round symbols) are in good agreement with those reported in [15] (square symbols). Then we used the semi-analytical formulation to calculate the mobility as a function of charge density for a uniform charge distribution (or uniform quasi-Fermi level, $E_{Fi} = E_{Fj}$) and same electric field ($F$=$10^3$V/cm) while using the transport energy ($E_t$) as a fitting parameter to match the Monte Carlo results.

The full line in Figure 1 is the result of such a fit and the inset to Figure 1 shows the resulting transport energy as a function of charge density. This procedure eliminated any arbitrariness in choosing the transport energy for the whole spectrum of carrier densities, and is consistent with the fact that the transport energy is known to be slightly dependent on the carrier density.[27] For completeness the dashed line shows the results with the transport energy being fixed at its value for low charge density.

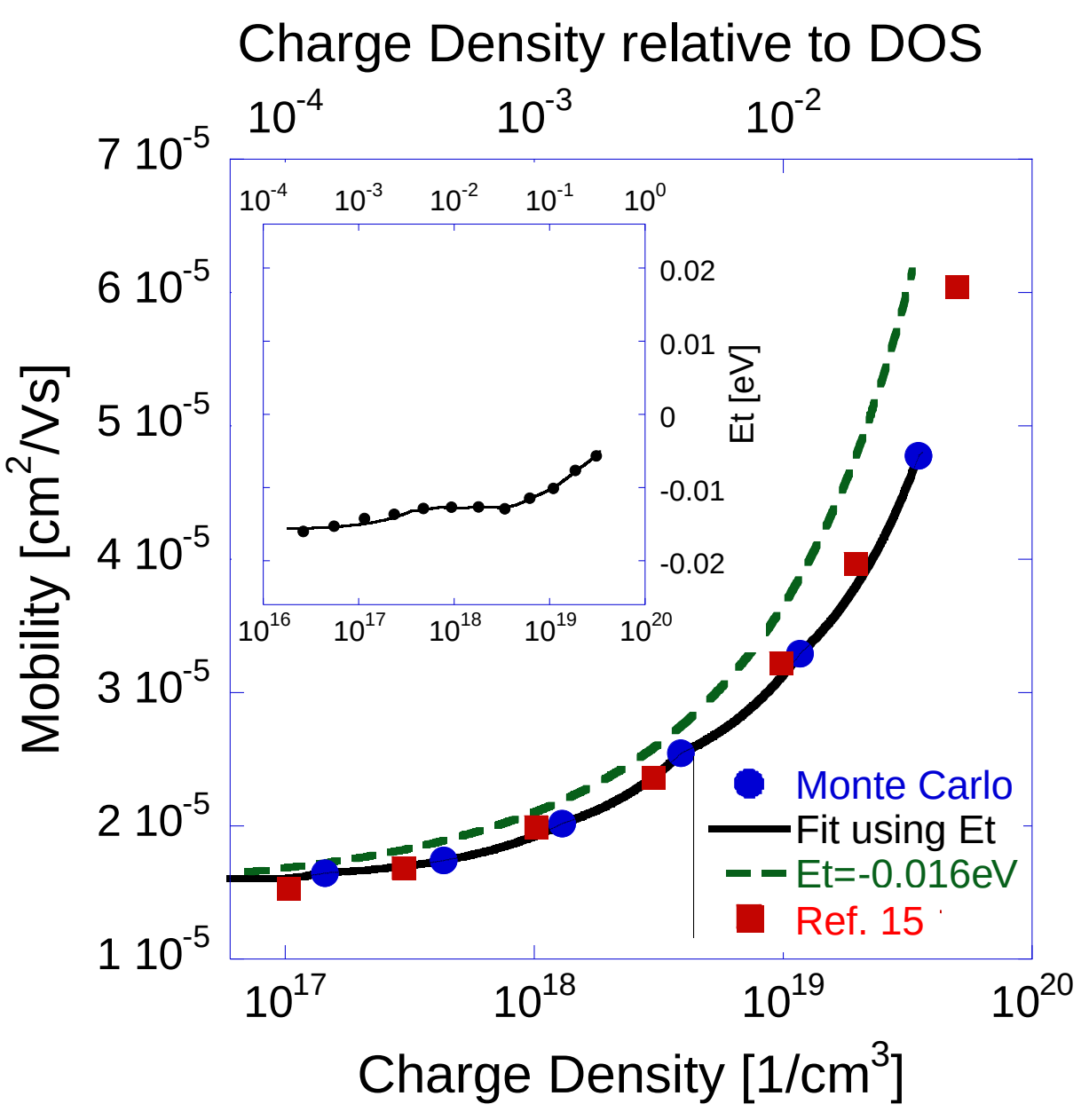

Figure 1. The mobility as a function of carrier density obtained via a homogenous carrier distribution Monte Carlo simulation (round symbols) under applied electric field of $F=10^3$[V/cm] and using a Gaussian DOS with standard deviation $\sigma$=3kT. Square symbols denote data taken from ref [15]. The full line is the mobility found by fitting the transport energy and the dashed line is for using transport energy found for low charge density to calculate the entire mobility range. The inset shows the fitted transport energy the value of which changes by less than kT.

***Charge transport evaluation*** - In order to verify that the procedure employed and presented in Figure 1 indeed ensures that in the present context the EMA based semi-analytic equation (2) is matched to the Monte Carlo simulations, we employed Monte Carlo simulations and solved equation (2) using several boundary conditions that would induce charge density gradients. In Figure 2a and in Figure 2b we set the charge density at x=0 at a relatively high value and at the opposite side we set it to zero. In Figure 2a the applied electric field is such that it would induce current in the same direction as the diffusion would and in Figure 2b it opposes the diffusion. In Figure 2c the boundary conditions are chosen to enforce equilibrium and hence the two ends are implemented as blocking contacts (zero steady state current). In this last calculation we set the total number of charge carriers as the initial condition and set the electric field in a given direction. In all three sub figures the symbols represent the Monte Carlo simulation results and the full lines are the results of the EMA based semi-analytic equation (2).

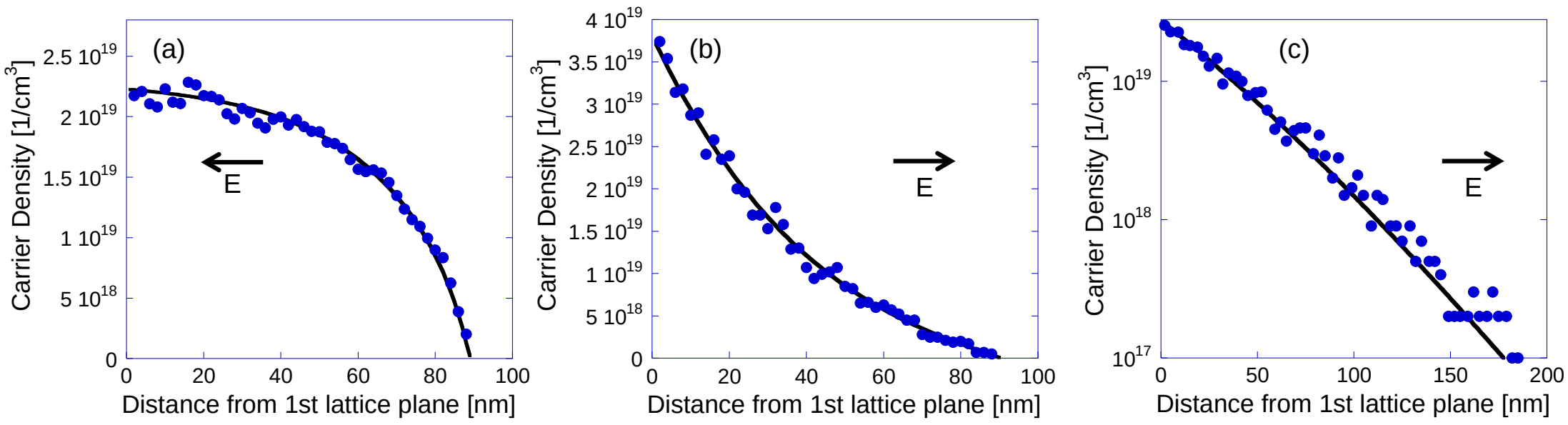

Figure 2. Charge density distribution calculated using the Monte Carlo simulation (symbols) and the EMA based semi-analytic equation (2) (full line). (a) and (b) present solutions for boundary conditions of high charge density at x=0 and zero at the other end. (c) Solutions for blocking boundary conditions (zero current) and a fixed total number of carriers under low electric field where the field force direction is indicated in each sub-figure.

Examining the three sub-figures of Figure 2 we note that both under equilibrium (zero-current) and under non-equilibrium conditions the two formulations agree and that this agreement holds both when the drift and diffusion currents are aligned and when they are opposing. It is worth noting that the boundary conditions were chosen such that between the three sub-figures a wide range of charge density gradients is covered.

Having established the correlation between the Monte Carlo simulation and the semi-analytical formula for the hopping rates (equation (3)) we use the latter to calculate the mobility, as defined for the drift-diffusion equation (2), under varying charge density gradients. Figure 3a presents calculations of the mobility as a function of the charge density assuming a local gradient in the quasi-Fermi level ($E_F$). The value next to each line depicts the difference in eV of $E_F$ between two successive planes separated by 1nm. A positive energy gradient is defined such that the carrier density gradient and applied field are enforcing current in the same direction. We note that when using the relation subsequent from equation (2), the resulting mobility is a function of both the charge density and the charge density gradient.

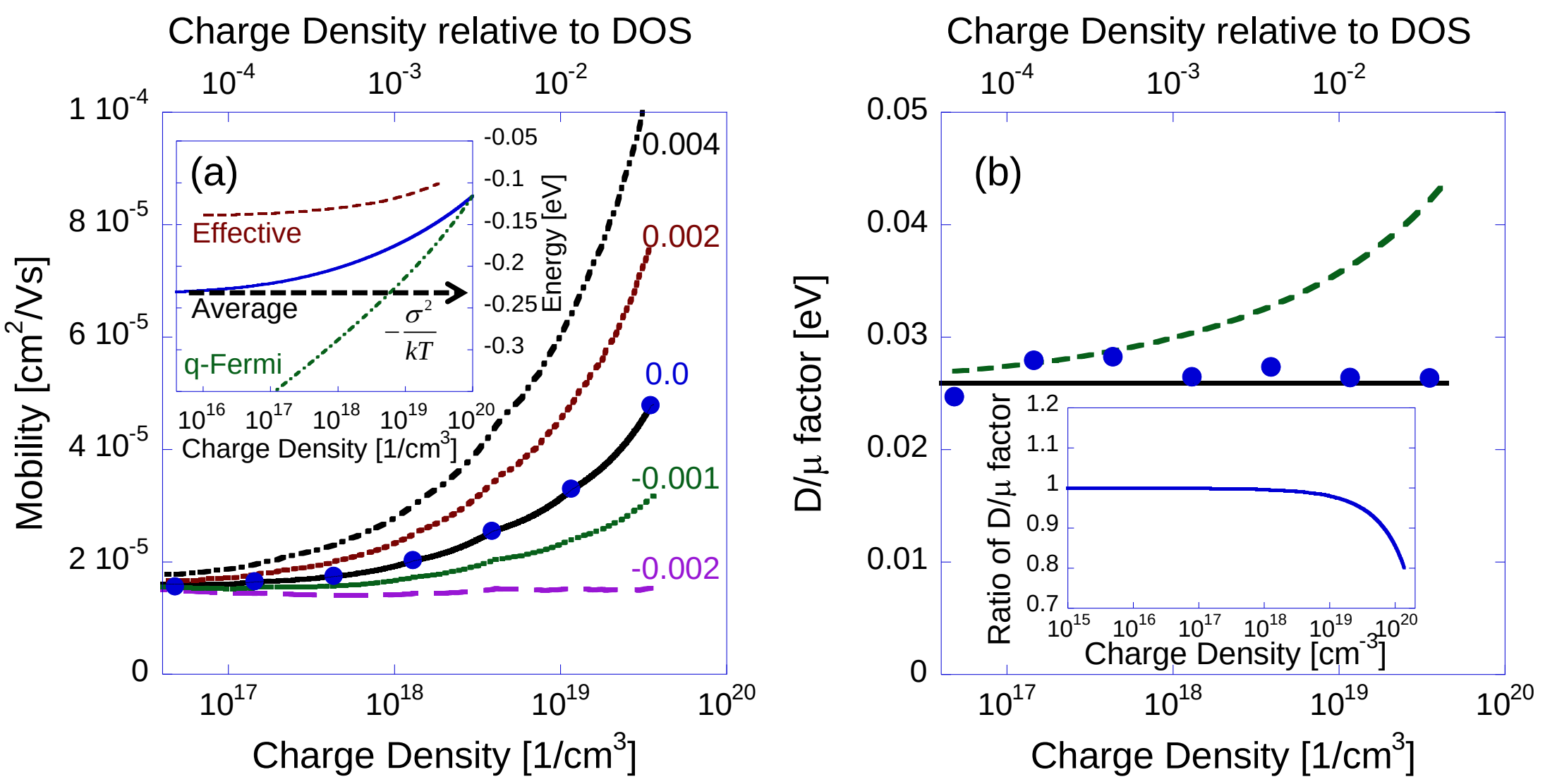

Figure 3. (a) The mobility as a function of charge density calculated using the definition in equation (2). The symbols are results obtained using the Monte Carlo simulation (Figure 1) for the case of uniform charge density. The other lines were calculated assuming a gradient in the quasi-Fermi level and the difference, in eV, between two planes is marked next to each line. The inset shows as a function of the charge density the quasi-Fermi level (dashed-dotted line), the average energy (full line), and the effective energy (dashed line). See discussion section for details. (b) The Einstein relation calculated using the uniform density Monte Carlo simulation used also for Figure 1 (symbols) and the EMA based semi-analytic equation (2) (full line). The dashed line represents the standard generalized Einstein relation[11]. The inset shows the ratio between the relation $D/\mu$ calculated using our new formalism and the ratio derived using the generalized Einstein relation (see also equation (6)).

Figure 3b depicts another aspect of the system studied. The theory leading to the generalized Einstein relation [11, 28] predicts that for a degenerate semiconductor the ratio between the diffusion and drift coefficients would increase as a function of the charge density (dashed line in Figure 3b). However, the calculations (full line) based on the hopping rates (eq. (3)) predict that when the charge density gradient is zero this relation remains at its classical value of kT/q. The symbols in Figure 3b are calculations derived directly from the Monte Carlo simulations used to evaluate the mobility presented in Figure 1 following the procedure in [24]. Figure 3b strengthens the notion developed in Figure 3a that in a degenerate hopping system, and within the framework of equation (1), the charge density gradient seem to affect both the diffusion and mobility processes.

***Discussion and Conclusions*** – In the earlier days of evaluating the mobility in disordered organic semiconductors the aspect of charge density and its effect on the mobility was given little to no attention. In the past decade it has become evident that the charge density plays a very important role and as indicated in the introduction, several models and calculations have been presented. Accounting for these semiconductors being degenerate under almost all practical experimental conditions we have added another attribute affecting the mobility – the charge density gradient. In order to gain better physical insight, to the Monte Carlo simulations we coupled the EMA based semi-analytic equation (2) as the latter is much handier in analyzing the data.

To understand the dependence on the charge density gradient, shown in Figure 3 we plot in the inset to Figure 3a the position of the quasi-Fermi level (dashed-dotted line) and the average energy of the charges (full line) as a function of the charge density. We note that at low charge densities the average energy per charge is constant but once the material becomes degenerate this average energy increases as a function of charge density. A higher average energy implies a higher average energy per charge carrier, thus in the presence of a charge density gradient there would also be a gradient in the energy per charge carrier and hence one should expect an energy flow from high to low concentrations.

To check if the above effect may explain the results shown in Figure 3a and Figure 3b, we rely on the framework of the EMA that uses the concept of transport-energy (equation (3)) where the leading terms in the rates at and against the field direction would be $W_e^+ = \nu_0 \exp\left(-\frac{2a}{b}\right)\exp\left(-\frac{\left(E_t + qFa - \bar{E}_i\right)}{K_B T}\right)$ and

$W_e^- = \nu_0 \exp\left(-\frac{2a}{b}\right)\exp\left(-\frac{\left(E_t - \bar{E}_j\right)}{K_B T}\right)$, respectively. Where $\bar{E}$ serves as the effective energy from which carriers are hopping. If the hopping rate was independent of the charge energy than the effective energy $\left(\bar{E}\right)$ would be equal to the average energy $\left(E_{avg}\right)$, plotted as a full line in the inset to Figure 3a. However, since hopping from higher energies to the transport energy is faster than from lower energies, the hopping rate of carriers in the high energy tail is faster resulting in $\bar{E} > E_{avg}$ . The dashed lines in the inset to Figure 3a show $\bar{E}$ calculated for the hopping rate used in this paper (i.e. Miller Abraham). Using this notation, the current flowing from i to j or from x to x+dx can be expressed as:

$$J = -qa\nu_0 \exp\left(-\frac{2a}{b}\right)\cdot\left[n(x)\exp\left(-\frac{E_t - (\bar{E}_i - qFa)}{K_B T}\right) - (n(x) + \Delta n(x))\exp\left(-\frac{E_t - \bar{E}_j}{K_B T}\right)\right]$$

Assuming that the field is not high and that the effective energy changes slowly

$\left(i.e.\ \left[\bar{E}_i - \bar{E}_j\right], qFa << K_B T\right)$ the above expression can be approximated to give:

$$J \overset{1st.\ order}{\approx} qav_0 \exp\left(-\frac{2a}{b}\right)\left\{-n(x)\exp\left(-\frac{E_t - \bar{E}_i}{K_B T}\right)\left[1 - \frac{Fa}{K_B T/q} - \left(1 + \left(\frac{\bar{E}_j - \bar{E}_i}{K_B T}\right)\right)\right] + \Delta n(x)\exp\left(-\frac{E_t - \bar{E}_j}{K_B T}\right)\right\}$$

Using the notion that the homogenous hopping rate in the absence of electric field can be expressed as: $W(n_i) = v_0 \exp\left(-\frac{2a}{b}\right)\exp\left(-\frac{E_t - \bar{E}_i}{K_B T}\right)$ and that the diffusion coefficient for the case of uniform or low charge density can be expressed as: $D_2(n_i) = a^2 W(n_i)$ one can arrive at:

$$J \overset{1st.\ order}{\approx} q\frac{n(x)D_2(n_i)}{K_B T}\left[qF + \frac{\bar{E}_j - \bar{E}_i}{a}\right] + q\frac{\Delta n(x)}{a}D_2(n_i)$$

Accounting for the result shown Figure 3b $\left(\frac{D_2}{K_B T/q} = \mu_2\right)$ and for $(a = \Delta x)$:

(4) $J = qn(x)\mu_2(x)F(x) + n(x)\mu_2(x)\frac{d\bar{E}}{dx} + q\frac{dn}{dx}D_2(x)$

Where $\mu_2(x)$ is the mobility in the case of uniform or low carrier distributions. To examine the generality of equation (4) we add: $d\bar{E} = \frac{\partial \bar{E}}{\partial n}dn + \frac{\partial \bar{E}}{\partial T}dT$ .

We note that if the system is non-degenerate $\frac{\partial \bar{E}}{\partial n} = 0$ and equation (4) reduces to the form of the energy balance equation, as derived for non-degenerate semiconductors (see eq. 44 in [29] & chapter 3.9 in [17] where $\bar{E} = \frac{3}{2}K_B T$ is used to derive the final equation). Also, if the system is non-degenerate ($\frac{\partial \bar{E}}{\partial n} = 0$) and there are no thermal gradients (dT=0) then $\frac{d\bar{E}}{dx} = 0$ and equation (4) reduces to equation (1). If the system is degenerate, however, $\frac{d\bar{E}}{dx}$ is non-zero even in the absence of thermal gradients. Due to the above, we consider that including this effect into the current equation is more general than using equation (1) and defining the coefficient D through the generalized Einstein relation, as in equation (5):

$$(5)\quad J = qD_3\frac{dn}{dx} + qn\mu_3 F;\;\; D_3 = \frac{KT}{q}\frac{n}{\partial n/\partial E_F}\mu_3(x)$$

This last statement actually requires some additional discussion. We consider equation (4) to be the most explicit representation with $D_2$ and $\mu_2$ standing for the intuitive, low density limit, diffusion and drift processes. Here, the effect of charge density gradient is encapsulated in the middle term $\left(\frac{d\bar{E}}{dx}\right)$. In equation (2) $\mu_1$ is defined such that it includes the effect due to energy transport $\left(\frac{d\bar{E}}{dx}\right)$. In equation (5) it is the diffusion coefficient that is chosen to encapsulate the energy transport term. If one were to compare the proposed formulation with that of the generalized Einstein relation, the energy flux term would need to be included as part of the diffusion coefficient thus ensuring that the mobility in both approaches is defined in the same way, i.e. for uniform density distributions, meaning ( $\mu_3 = \mu_2$ ). Then, if the use of the generalized Einstein relation in equation (5) is to result in an equation that is equivalent to equation (4) than the following should hold:

$$(6)\quad \left(n(x)\frac{1}{q}\frac{d\bar{E}}{dn} + \frac{KT}{q}\right)\mu(x) \stackrel{?}{=} \left(\frac{KT}{q}\frac{n}{\partial n/\partial E_F}\right)\mu(x)$$

While we are not able to analytically test this equality the inset to Figure 3b shows the ratio of the left to the right sides of equation (6). As this sub figure shows, the use of the generalized Einstein is a good enough approximation up to densities that are ~$10^{-2}$ of the total DOS. Another point arising from equation (4) is that it is common to state that if the quasi-Fermi level is constant than there would be no currents. However, it is clear that this statement is true only if there are no temperature gradients in the sample. In fact the above statement breaks whenever $\frac{d\bar{E}}{dx} \neq 0$ and hence it does not hold for the degenerate case either.

To conclude, having established that the hopping rate can be expressed using the transport energy concept we were able to show (equation (4)) that in a disordered semiconductor (i.e. where $\frac{\partial \bar{E}}{\partial n} <> 0$ ) there is an energy flow $n(x)\mu_2(x)\frac{\partial \bar{E}}{\partial x}$ that is non-negligible and that it accounts for the “anomalies” described in Figure 3a. We also note that although equation (4) was derived in the context of hopping transport it can also be derived from the basic Boltzmann equation ensuring that the expression used for $\bar{E}$ accounts for the fact that for the degenerate case $\bar{E} = \bar{E}(n,T)$ . Namely, if one adds also an equation to follow the energy balance (energy flow) one can also model the carrier heating phenomena[8] with the transport equations.

**Acknowledgments –** This research was partially supported by the Ollendorff Minerva Center. We acknowledge discussions with S. D. Baranovskii and P.A. Bobbert.

## References

1 H. Scher and E. M. Montroll, Phys. Rev. B **12**, 2455 (1975).

2 N. Rappaport, Y. Preezant, and N. Tessler, Phys. Rev. B **76**, 235323 (2007).

3 W. D. Gill, J. Appl. Phys. **43**, 5033 (1972).

4 H. Bassler, G. Schonherr, M. Abkowitz, and D. M. Pai, Phys. Rev. B - Cond. Matt. **26**, 3105 (1982).

5 B. I. Shklovskii, E. I. Levin, H. Fritzsche, and S. D. Baranovskii, in *Transport, Correlation and Structural Defects*, edited by H. Fritzsche (World Scientific, Singapore, 1990), p. 161.

6 M. Van der Auweraer, F. C. Deschryver, P. M. Borsenberger, and H. Bassler, Advanced Materials **6**, 199 (1994).

7 Z. G. Yu, D. L. Smith, A. Saxena, R. L. Martin, and A. R. Bishop, Phys. Rev. B **63**, 085202 (2001).

8 Y. Preezant and N. Tessler, Phys. Rev. B **74**, 235202 (2006).

9 P. E. Parris and B. D. Bookout, Phys. Rev. B **53**, 629 (1996).

10 V. Ambegaokar, B. I. Halperin, and J. S. Langer, Phys. Rev. B **4**, 2612 (1971).

11 Y. Roichman and N. Tessler, Applied Physics Letters **80**, 1948 (2002).

12 S. D. Baranovskii, I. P. Zvyagin, H. Cordes, S. Yamasaki, and P. Thomas, Physica Status Solidi B-Basic Research **230**, 281 (2002).

13 S. Shaked, S. Tal, Y. Roichman, A. Razin, S. Xiao, Y. Eichen, and N. Tessler, Advanced Materials **15**, 913 (2003).

14 V. I. Arkhipov, P. Heremans, E. V. Emelianova, G. J. Adriaenssens, and H. Bassler, Applied Physics Letters **82**, 3245 (2003).

15 W. F. Pasveer, J. Cottaar, C. Tanase, R. Coehoorn, P. A. Bobbert, P. W. M. Blom, D. M. de Leeuw, and M. A. J. Michels, Physical Review Letters **94**, 206601 (2005).

16 R. Coehoorn, W. F. Pasveer, P. A. Bobbert, and M. A. J. Michels, Phys. Rev. B **72**, 155206 (2005).

17 J. Piprek, *Semiconductor Optoelectronic Devices: Introduction to Physics and Simulation* (Academic Press, 2003).

18 N. Tessler, Y. Preezant, N. Rappaport, and Y. Roichman, Advanced Materials **21**, 2741 (2009).

19 Y. Vaynzof, Y. Preezant, and N. Tessler, Journal of Applied Physics **106**, 6 (2009).

20 I. N. Hulea, H. B. Brom, A. J. Houtepen, D. Vanmaekelbergh, J. J. Kelly, and E. A. Meulenkamp, Phys. Rev. Lett. **93**, 166601 (2004).

21 O. Tal, Y. Rosenwaks, Y. Preezant, N. Tessler, C. K. Chan, and A. Kahn, Physical Review Letters **95**, 256405 (2005).

22 J. O. Oelerich, D. Huemmer, and S. D. Baranovskii, Physical Review Letters **108**, 226403 (2012).

23 N. Rappaport, O. Solomesch, and N. Tessler, Journal of Applied Physics **99**, 064507 (2006).

24 A. V. Nenashev, F. Jansson, S. D. Baranovskii, Ouml, R. sterbacka, A. V. Dvurechenskii, and F. Gebhard, Phys. Rev. B **81**, 115204 (2010).

25 I. I. Fishchuk, D. Hertel, H. Bässler, and A. K. Kadashchuk, Phys. Rev. B **65**, 125201 (2002).

[26] I. I. Fishchuk, V. I. Arkhipov, A. Kadashchuk, P. Heremans, and H. Bässler, Phys. Rev. B **76**, 045210 (2007).
[27] J. O. Oelerich, D. Huemmer, M. Weseloh, and S. D. Baranovskii, Applied Physics Letters **97**, 143302 (2010).
[28] N. W. Ashcroft and N. D. Mermin, *Solid State Physics* (HOLT, RINEHART AND WINSTON, New York, 1988).
[29] R. Stratton, Ieee Transactions on Electron Devices **ED19**, 1288 (1972).